# Experimental realization of an intrinsic magnetic topological insulator

Yan Gong[1], Jingwen Guo[1], Jiaheng Li[1], Kejing Zhu[1], Menghan Liao[1], Xiaozhi Liu[2], Qinghua Zhang[2], Lin Gu[2], Lin Tang[1], Xiao Feng[1], Ding Zhang[1,5], Wei Li[1,5], Canli Song[1,5], Lili Wang[1,5], Pu Yu[1,5], Xi Chen[1,5], Yayu Wang[1,5], Hong Yao[3,5], Wenhui Duan[1,5], Yong Xu[1,6*], Shou-Cheng Zhang[4], Xucun Ma[1,5], Qi-Kun Xue[1,5*], Ke He[1,5*]

[1]*State Key Laboratory of Low Dimensional Quantum Physics, Department of Physics, Tsinghua University, Beijing 100084, China*

[2]*Beijing National Laboratory for Condensed Matter Physics, Institute of Physics, Chinese Academy of Sciences, Beijing 100190, China*

[3]*Institute for Advanced Study, Tsinghua University, Beijing 100084, China*

[4]*Stanford Center for Topological Quantum Physics, Department of Physics, Stanford University, Stanford, California 94305-4045, USA*

[5]*Collaborative Innovation Center of Quantum Matter, Beijing 100084, People's Republic of China*

[6]*RIKEN Center for Emergent Matter Science (CEMS), Wako, Saitama 351-0198, Japan*

Emails:  yongxu@mail.tsinghua.edu.cn
  qkxue@mail.tsinghua.edu.cn
  kehe@tsinghua.edu.cn

**Intrinsic magnetic topological insulator (TI) is a stoichiometric magnetic compound possessing both inherent magnetic order and topological electronic states. Such a material can provide a shortcut to various novel topological quantum effects but remains elusive experimentally so far. Here, we report the experimental realization of high-quality thin films of an intrinsic magnetic TI—MnBi$_2$Te$_4$—by alternate growth of a Bi$_2$Te$_3$ quintuple-layer and a MnTe**

**bilayer with molecular beam epitaxy. The material shows the archetypical Dirac surface states in angle-resolved photoemission spectroscopy and is demonstrated to be an antiferromagnetic topological insulator with ferromagnetic surfaces by magnetic and transport measurements as well as first-principles calculations. The unique magnetic and topological electronic structures and their interplays enable the material to embody rich quantum phases such as quantum anomalous Hall insulators and axion insulators in a well-controlled way.**

A topological insulator (TI) is non-magnetic, carrying gapless surface electronic states topologically protected by the time-reversal symmetry (TRS) (*1*, *2*). Many exotic quantum effects predicted for TIs, however, need the TRS to be broken by acquired magnetic order (*3*). A remarkable example is the quantum anomalous Hall (QAH) effect, a zero-magnetic-field quantum Hall effect that had been sought after for over two decades until it was observed in a magnetic TI with ferromagnetic (FM) order induced by magnetic dopants (*3-6*). The experimental realization of the QAH effect paved the road for hunting many other novel quantum effects in TRS-broken TIs, for example topological magnetoelectric (TME) effects and chiral Majorana modes (*3*, *8*, *9*). However, magnetically doped TIs are notorious "dirty" materials for experimental studies: the randomly distributed magnetic impurities induce strong inhomogeneity in the electronic structure and magnetic properties, and the sample quality is sensitive to the details of the molecular beam epitaxy (MBE) growth conditions (*10-12*). Such a complicated system is often a nightmare for some delicate experiments such as those on chiral Majorana modes and topological quantum computation, and the strong inhomogeneity is believed to contribute to the extremely low temperature (usually <100 mK) required by the QAH effect (*13*). An ideal magnetic TI is an intrinsic one, namely a stoichiometric compound with orderly arranged and exchange-coupled magnetic atoms which features a magnetically ordered ground state, but becomes a TI when the TRS recovers above the magnetic ordering temperature. A thin film of such an intrinsic magnetic TI could be a congenital QAH insulator with homogeneous electronic and magnetic properties, and presumably higher QAH working temperature. Yet few experimental progresses were achieved in this direction in spite of several interesting theoretical proposals raised in the past years (*14-16*).

Some stoichiometric ternary tetradymite compounds, which can be considered as variants of well-studied $Bi_2Te_3$ family 3D TIs, have been found to be also 3D TIs (*17*). A simplest system is $XB_2T_4$ where X is Pb, Sn or Ge, B is Bi or Sb, and T is Te or Se. Such a compound is a layered material with each septuple-layer (SL) composed of single atomic sheets stacking in the sequence T-B-T-X-T-B-T. If X were a magnetic element, there would be chance that $XB_2T_4$ is an intrinsic magnetic TI. A few works observed $MnBi_2Te(Se)_4$ in multi-crystalline samples, or as the second phase or surface layer of $Bi_2Te(Se)_3$, without figuring out their topological electronic properties (*18-20*). Interestingly, a SL of $MnBi_2Te(Se)_4$ on $Bi_2Te(Se)_3$ was reported to be able to open a large magnetic gap at the topological surface states of the latter (*20, 21*).

In this study, we found that high-quality $MnBi_2Te_4$ films can be fabricated in a SL-by-SL manner by alternate growth of 1 quintuple-layer (QL) of $Bi_2Te_3$ and 1 bilayer (BL) of MnTe with MBE. Amazingly, $MnBi_2Te_4$ films with the thickness $d \geqslant$ 2 SL show Dirac-type surface states—a characteristic of a 3D TI. Low temperature magnetic and transport measurements as well as first-principles calculations demonstrate that $MnBi_2Te_4$ is an intrinsic antiferromagnetic (AFM) TI, composed of ferromagnetic SLs with a perpendicular easy axis which are coupled antiferromagnetically between neighboring SLs. Remarkably, a thin film of such an AFM TI thin film with FM surfaces is expected to be an intrinsic QAH insulator or axion insulator depending on the film thickness.

To prepare a $MnBi_2Te_4$ film, we first grow a 1 QL $Bi_2Te_3$ film on a Si(111) or $SrTiO_3$(111) substrate (*22, 23*). Mn and Te are then co-evaporated onto $Bi_2Te_3$ surface with the coverage corresponding to a MnTe BL with the sample kept at 200°C. Post-annealing at the same temperature for 10 minutes is carried out to improve the crystalline quality. This leads to the formation of a SL of $MnBi_2Te_4$ (see the schematic in Fig. 1A) (*20*), as experimentally proved and theoretically explained below. Then on the MnBi$_2$Te$_4$ surface, we grow another QL of Bi$_2$Te$_3$ which is followed by deposition of another BL of MnTe and post-annealing. By repeating this procedure, we can grow a MnBi$_2$Te$_4$ film SL by SL in a controlled way, in principle up to any desired thickness.

The MnBi$_2$Te$_4$ film shows sharp 1×1 reflection high-energy electron diffraction streaks (Fig. S1) indicating its flat surface morphology and high crystalline quality. The X-ray diffraction (XRD) pattern (Fig. 1B, taken from a 7 SL MnBi$_2$Te$_4$ film) exhibits a series of peaks (marked by blue arrows), most of which can neither be attributed to Bi$_2$Te$_3$ nor to MnTe. From the positions of these XRD peaks, we can estimate the spacing between the crystalline planes to be ~1.36 nm, very close to the inter-SL distance of bulk MnBi$_2$Te$_4$ (1.356 nm) predicted by our first-principles calculations.

High resolution scanning transmission electron microscopy (STEM) was used to characterize the real-space crystalline structure of a MnBi$_2$Te$_4$ film (5 SL). The high-angle annular dark field (HAADF) images (Figs. 1C and 1D) clearly show the characteristic SL structure of XB$_2$T$_4$ compounds, except for the region near the substrate where stack faults and dislocations are observed. Figure 1E displays the intensity profile along an atomic row across two SLs (cut 1 in Fig. 1C). One can see the atomic contrast varies a lot at different positions in a SL. The contrast of an atom in a HAADF-STEM image is directly related to its atomic number. The intensity distribution along a SL is thus well consistent with the Te-Bi-Te-Mn-Te-Bi-Te sequence. The electron energy lose spectroscopy (EELS) (Fig. 1F) reveals the Mn L$_{2,3}$ edges at ~645 eV. The intensity distribution curve of EELS at 645 eV (the pink line in Fig. 1F) taken along cut 2 in Fig. 1C shows a peak at the middle atom of each SL, which also agrees with the MnBi$_2$Te$_4$ structure.

In-situ angle-resolved photoemission spectroscopy (ARPES) was used to map the electronic energy band structure of the MBE-grown MnBi$_2$Te$_4$ films. Figures 2A-2D show the ARPES bandmaps of the MnBi$_2$Te$_4$ films with the thickness $d$ = 1, 2, 5, and

7 SL, respectively, with the sample temperature at ~ 25 K. The spectra were taken around Γ point along the M-Γ-M direction of the Brillouin zone. The spectra of the $d$ = 1 SL sample (Fig. 2A) shows a bandgap with Fermi level cutting the conduction band. The films with $d \geqslant 2$ SL all show similar band structures (Figs. 2B-2D). One can always observe a pair of energy bands with nearly linear band dispersion crossing at Γ point forming a Dirac cone. Figures 2E and 2F show the momentum distribution curves (MDCs) and the constant-energy contours of the 7 SL sample, respectively, which exhibits an archetypal Dirac-type energy bands. It is worth to note that the Dirac-type bands are quite different from the topological surface states of $Bi_2Te_3$ (*24, 25*). The band dispersion observed here is rather isotropic, as shown by the nearly circular constant-energy contours even at the energy far away from the Dirac point, which is distinct from the strongly warped $Bi_2Te_3$ topological surface states (*25, 26*). The Dirac point observed here is located right in the band gap, in contrast with the $Bi_2Te_3$ case where the Dirac point is below the valance band maximum. Moreover, the Fermi velocity near Dirac point is $5.5 \pm 0.5 \times 10^5$ m/s, obviously larger than that of $Bi_2Te_3$ surface states ($3.87$~$4.05 \times 10^5$ m/s in different directions) (*25*). Therefore the Dirac-type bands can only be attributed to $MnBi_2Te_4$, and, as demonstrated below, are also the topological surface states of a 3D TI.

The orderly and compactly arranged Mn atoms in $MnBi_2Te_4$ are expected to give rise to a long-range magnetic order at low temperature. Figure 3A displays the magnetization ($M$) — magnetic field ($H$) curves of a 7 SL $MnBi_2Te_4$ film measured with superconducting quantum interference device (SQUID) at different temperatures ($T$s). The linear diamagnetic background contributed by the substrate and capping layer has been subtracted (the raw data are shown in Fig. S2). The unit of $M$ is the magnetic moment ($\mu_B$) per in-plane unit cell (2D U.C.), i.e. the average magnetic moment of each Mn atom multiplied by the number of SLs. $H$ is applied perpendicular to the sample plane. With decreasing temperature, hysteresis appears in the *M-H* curves and grows rapidly, exhibiting a typical FM behavior. The Curie temperature ($T_C$) is 20 K according to the temperature ($T$) dependence of the remnant magnetization [$M_r = M(0\,T)$] shown in Fig. 3B. The *M-H* curve measured with in-plane magnetic field has much smaller hysteresis than the curve measured with perpendicular one (see Fig. 3A inset, which were taken from another 7 SL MnBi$_2$Te$_4$ sample). Therefore the magnetic easy axis is along the *c* direction [perpendicular to the (001) plane]. Estimated from the saturation magnetization $M_s = 8\,\mu_B$/2D U.C., the Mn atomic magnetic moment is about 1.14 $\mu_B$ which is much smaller than 5 $\mu_B$ expected for Mn$^{2+}$ ions. It suggests that Mn$^{2+}$ ions in the material may have a more complex magnetic structure than a simple uniform ferromagnetic configuration.

The ferromagnetism of 7 SL MnBi$_2$Te$_4$ film is also demonstrated by Hall measurements. Figure 3D displays the Hall resistance ($R_{yx}$) vs. *H* curves of a 7 SL film grown on SrTiO$_3$(111) substrate measured at 1.6 K under different gate-voltages ($V_g$s). The SrTiO$_3$ substrate is used as the gate dielectric for its huge dielectric constant (~20000) at low temperature (*27*). The curves exhibit hysteresis loops of the anomalous Hall effect (AHE) with a linear background contributed by the ordinary Hall effect (OHE). The slope of the OHE background reveals that the sample is electron-doped with the electron density $n_e \sim 1.1 \times 10^{13}$ cm$^{-2}$, which basically agrees with $n_e \sim 8 \times 10^{12}$ cm$^{-2}$ derived from the Fermi wavevector ($k_F \sim 0.07$Å$^{-1}$) of the ARPES-measured Dirac-type band. The hysteresis loops of the AHE confirm the ferromagnetism of the film with perpendicular magnetic anisotropy. The $T_C$ obtained from the $R_{yx}$-*T* curve is similar to that given by SQUID data (Fig. 2B). The $H_c$ of the $R_{yx}$-*H* hysteresis loops is however larger than that of the *M-H* loops. Tuning the chemical potential of the film by applying different $V_g$s, we observe obvious change in the anomalous Hall resistance. The sensitivity of the AHE to the chemical potential suggests that the AHE is mainly contributed by the Berry curvature of the energy bands induced by intrinsic magnetism of the material instead of magnetic impurities or clusters (*28*).

Noticeably, 6 SL MnBi$_2$Te$_4$ film shows different magnetic properties from 7 SL one. As shown in Fig. 3C, the hysteresis ($M_r$ and $H_c$) in the *M-H* curve of a 6 SL film is rather small even at 3 K, and $M_s$ decreases slowly with increasing temperature.

Clearly the film is not dominated by long-range FM order. The *M-H* curves of the MnBi$_2$Te$_4$ films from 4 SL to 9 SL are displayed in Fig. 3E which will be analyzed below based on our theoretical results.

Next we discuss the structure, magnetism and topological electronic properties of MnBi$_2$Te$_4$ with the above experimental observations and our first-principles calculation results. To understand the mechanism for the formation of MnBi$_2$Te$_4$, we calculated the energies of a MnTe BL adsorbed on a Bi$_2$Te$_3$ QL (Fig. 4A left) and a MnBi$_2$Te$_4$ SL (Fig. 4A right). The calculations show that the latter one has 0.51 eV/unit lower total energy and is thus energetically more stable. The result is easy to understand in terms of valence states. By assuming Te$^{2-}$, the former structure gives unstable valence states of Mn$^{3+}$ and Bi$^{2+}$ which tend to change into more stable Mn$^{2+}$ and Bi$^{3+}$ by swapping their positions. The atom-swapping induced stabilization thus explains the spontaneous formation of MnBi$_2$Te$_4$ with a MnTe BL grown on Bi$_2$Te$_3$.

We calculated the energies of different magnetic configurations of MnBi$_2$Te$_4$ (Fig. S3) (*23*). It was found that the most stable magnetic structure is FM coupling in each SL and AFM coupling between adjacent SLs (i.e. *A*-type AFM), whose easy axis is out-of-plane (Fig. 4B). In MnBi$_2$Te$_4$, Mn atoms are located at the center of slightly distorted octahedrons that are formed by neighboring Te atoms. The FM intralayer coupling induced by Mn-Te-Mn superexchange interactions is significantly stronger than the AFM interlayer coupling built by weaker Mn-Te ⋯ Te-Mn super-superexchange interactions. Similar *A*-type AFM states were predicted to exist in other magnetic XB$_2$T$_4$ compounds (*29*).

Figure 4C shows the calculated band structure of a 7 SL MnBi$_2$Te$_4$ film. We can observe Dirac-like energy bands around Γ point, which basically agrees with the ARPES data, expect for a gap (~ 52 meV) at the Dirac point. All the films above 4 SL show similar band feature with nearly identical gap values at the Dirac point, implying that the gapped Dirac cone is an intrinsic surface feature of the material. Purposely tuning down the SOC strength in calculations, the gap first decreases to zero and then increases (inset of Fig. 4C), which suggests a topological phase transition and thus the topologically non-trivial nature of the gap. Actually our calculations on the system reveal that bulk MnBi$_2$Te$_4$ is a 3D AFM TI with Dirac-like surface states that are gapped by the FM (001) surfaces with out-of-plane magnetization (*29*, *30*).

As illustrated in Fig. 4D and confirmed numerically, the gapped surface states can be described by an effective Hamiltonian $H(\mathbf{k}) = (\sigma_x k_y - \sigma_y k_x) + m_z \sigma_z$, where $\sigma$ is the Pauli matrix with $\sigma_z = \pm 1$ referring to spin up and down, $m_z$ is the surface exchange field (*2*, *3*). For films thicker than 1 SL, hybridizations between top and bottom surfaces are negligible. Thus, their topological electronic properties are determined by the two isolated surfaces, which have the same (opposite) $m_z$ for odd (even) number of SLs and half-integer quantized Hall conductance of $e^2/2h$ or $-e^2/2h$ depending on the sign of $m_z$. Therefore, odd-SL MnBi$_2$Te$_4$ films are intrinsic QAH insulators with Chern number $C = 1$; meanwhile even-SL films are intrinsic axion insulators ($C = 0$) that behave like ordinary insulators in dc measurements but can show topological magnetoelectric effects in ac measurements (*3*). However, when the TRS is recovered above $T_C$, the exchange splitting of the bands gets vanished while the SOC-induced topological band inversion remains unaffected. MnBi$_2$Te$_4$ thus becomes a 3D TI showing gapless topological surface states which are exactly the band structure observed in the ARPES measurements performed at 25 K (above $T_C$).

The theoretically predicted magnetic configuration of MnBi$_2$Te$_4$ (Fig. 4B) is supported by our magnetic measurements. For an odd-SL AFM MnBi$_2$Te$_4$ film, whatever the exact thickness, the net magnetic moment is only of 1 SL. It explains why the atomic magnetic moment of Mn estimated from the 7 SL MnBi$_2$Te$_4$ film (1.14 $\mu_B$) is much smaller than 5 $\mu_B$. The measured $M_s$ = 8 $\mu_B$ per 2D U.C. may have contributions from both the FM surfaces (supposed to be 5 $\mu_B$) and the AFM bulk which can give magnetic signals via canting or disorder. With the AFM arrangement of neighboring FM SLs, MnBi$_2$Te$_4$ films are expected to show oscillation in its magnetic properties as the thickness changes between even and odd SLs. We indeed observed even-odd oscillation in their magnetic properties as shown in Figs. 3E and

3F. The remnant magnetization ($M_r$), which characterizes long range ferromagnetic order, is larger in odd-SL films than in even-SL ones. $H_c$ shows similar oscillation below 7 SL, but increases monotonously in thicker films. It is because in an AFM film with FM surfaces, the Zeeman energy in magnetic field ($E_z$) is only contributed by the FM surfaces and thus invariant with film thickness, while the magnetocrystalline anisotropy energy ($E_{MCA}$), which is contributed by the whole film, increases with thickness and thus becomes more difficult to be overcome by $E_z$. Besides, as shown in the 6 SL film (Fig. 3C) and other even-SL films, $M_s$ in less sensitive to temperature than in odd-SL films. For a comparison, the differences between the $M$-$H$ curves measured at 3 K and those measured above $T_C$ are displayed in the bottom column of Fig. 3E, which shows a clear even-odd oscillation (Fig. 3F). A rapid increase of $M_s$ with decreasing temperature below $T_C$ is typical of ferromagnetic order. The magnetic signal from AFM canting, on the other hand, decreases or keeps nearly constant with decreasing temperature. So the odd-SL films obviously have more FM features.

The large inter-SL distance (~1.36 nm) is expected to give a weak AFM coupling between neighboring SLs which can be aligned into FM configuration in a magnetic field of several tesla (*31*). We carried out a Hall measurement of a 7 SL MnBi$_2$Te$_4$ film with $H$ up to 9 T. As shown in Fig. 3G (the linear background of the OHE has been subtracted from the $R_{yx}$-$H$ loop), besides a small hysteresis loop at low field contributed by the FM surfaces, $R_{yx}$ resumes growing above ~ 2 T and is saturated at a higher plateau above 5 T. The phenomenon is typical of a layered magnetic material and presumably results from an AFM-to-FM transition (see the schematic magnetic configuration shown by the blue arrows in Fig. 3G). The FM configuration may drive the system into a magnetic Weyl semimetal phase (*29*, *30*).

In spite of the above evidences for an *A*-type AFM order of MnBi$_2$Te$_4$, there are still some observations which we have not yet fully understood. For example, the even-SL films show larger $M_s$ than odd-SL ones above $T_C$, which is particularly clear in comparing the 6 SL (Fig. 3C) and 7 SL (Fig. 3A) data at 30 K. We also notice that overall $M_s$ shows a maximum around 6 SL and 7 SL at 3 K, regardless of even- or odd-SLs. Another confusion is that the magnetic properties revealed by Hall effect measurements are not fully consistent with those revealed by magnetization measurements: $R_{yx}$-$H$ loops always show larger $H_c$ than $M$-$H$ loops, and oscillatory behaviors are barely observed in the AHE data of the films of different thicknesses. These phenomena should result from the interplays between the complex magnetic structures and topological electronic properties of the unique layered magnetic material and require a comprehensive study combing various techniques to clarify (*31*, *32*). Besides, we found that MnBi$_2$Te$_4$ films are relatively easy to decay at ambient condition: $M_s$ of a sample decreases significantly after it is exposed in air for couple of days. This may also complicate the magnetization and magneto-transport measurement results. Finding an effective way to protect the material is crucial for the experimental investigations on this system and for the explorations of the exotic topological quantum effects in it.

**Acknowledgments:** The authors thank Wanjun Jiang and Jing Wang for stimulating discussions. We are grateful to the National Science Foundation of China, Ministry of Science and Technology of China, and the Beijing Advanced Innovation Center for Future Chip (ICFC) for financial support.

**Figures Captions**

**Fig. 1. MBE growth and structural characterizations of MnBi$_2$Te$_4$ films.** (**A**) Schematic illustrations of the MBE growth mechanism of 1 septuple layer (SL) MnBi$_2$Te$_4$ thin film. (**B**) XRD pattern of a MnBi$_2$Te$_4$ (MBT) film grown on Si(111). (**C**) Cross-sectional HAADF-STEM image of a 5 SL MnBi$_2$Te$_4$ film grown on a Si (111) substrate. (**D**) Zoom-in view of (C) with the structural model of MnBi$_2$Te$_4$. (**E**) Intensity distribution of HAADF-STEM along cut 1 in (C). (**E**) EELS spectra mapping along cut 2 in (C). The pink curve shows the intensity distribution of the Mn L$_{2,3}$-edge along cut 2 in (C).

**Fig. 2. Energy band structures of MnBi$_2$Te$_4$ films measured by ARPES.** (**A-C**) ARPES spectra of 1, 2, 5, and 7 SL MnBi$_2$Te$_4$ films measured near the Γ point, along the M-Γ-M direction. (**D**) Momentum distribution curves (MDCs) of the 7 SL film from E$_F$ to −0.38 eV. The red triangles indicate the peak positions. (**E**) Constant energy contours of the 7 SL film at different energies. All the ARPES data were taken at 25 K.

**Fig. 3. Magnetic and magneto-transport properties of MnBi$_2$Te$_4$ films.** (**A**) Magnetization vs. magnetic field (*M-H*) curves of a 7 SL MnBi$_2$Te$_4$ film measured with SQUID at 3 K (red), 10 K (orange), 15 K (green), and 30 K (blue), respectively. *H* is perpendicular to the sample plane. The inset shows *M-H* curves measured with *H* perpendicular to (red) and in (blue) the sample plane (a different 7 SL MnBi$_2$Te$_4$ sample). (**B**) Temperature dependences of the remnant magnetization (*M*$_r$) and zero magnetic field Hall resistance (*R*$_{yx}^0$) of a 7 SL film, which give the Curie temperature (*T*$_C$). (**C**) *M-H* curves of a 6 SL MnBi$_2$Te$_4$ film measured with SQUID at 3 K (red), 10 K (orange), 15 K (green), and 30 K (blue), respectively. *H* is perpendicular to the sample plane. (**D**) *R*$_{yx}$-*H* curves measured at 1.6 K at different gate voltages. (**E**) *M-H* curves of 4, 5, 6, 7, 8, and 9 SL MnBi$_2$Te$_4$ films measured at 3 K and right above *T*$_C$ (top column) and the differences between the curves at the two temperatures (bottom

column). (**F**) Thickness dependences of $M_r$ at 3 K, $M_r$ difference at 3 K and above $T_C$ (top panel) and $H_C$ (bottom panel). (**G**) $R_{yx}$-$H$ curve of a 7 SL MnBi$_2$Te$_4$ film measured at 1.6 K with $H$ up to 9 T. The blue arrows indicate the magnetic configurations at different $H$. Each arrow represents the magnetization vector of a SL.

**Fig. 4. First-principles calculation results of MnBi$_2$Te$_4$.** (**A**) Lattice structures of a MnTe bilayer adsorbed on a Bi$_2$Te$_3$ quintuple layer (left) and a MnBi$_2$Te$_4$ SL (right). Valence states of atoms were labelled by assuming -2 for Te. Atom swapping between Mn and Bi results in stable valence states, thus stabilizing the whole structure. (**B**) Atomic structure of layered MnBi$_2$Te$_4$, whose magnetic states are ferromagnetic within each SL and antiferromagnetic between adjacent SLs. Insets show Te-formed octahedrons together with center Mn. (**C**) Band structure of a 7-SL MnBi$_2$Te$_4$ film, which is an intrinsic QAH insulator (band gap ~52 meV), as proved the dependence of band gap on the strength of SOC (inset). (**D**) Schematic band structure of MnBi$_2$Te$_4$ (001) surface states, showing a gapped Dirac cone with spin-momentum locking. The energy gap is opened by the surface exchange field ($m_z$), which gets vanished when paramagnetic states are formed at high temperatures.

# Figure 1

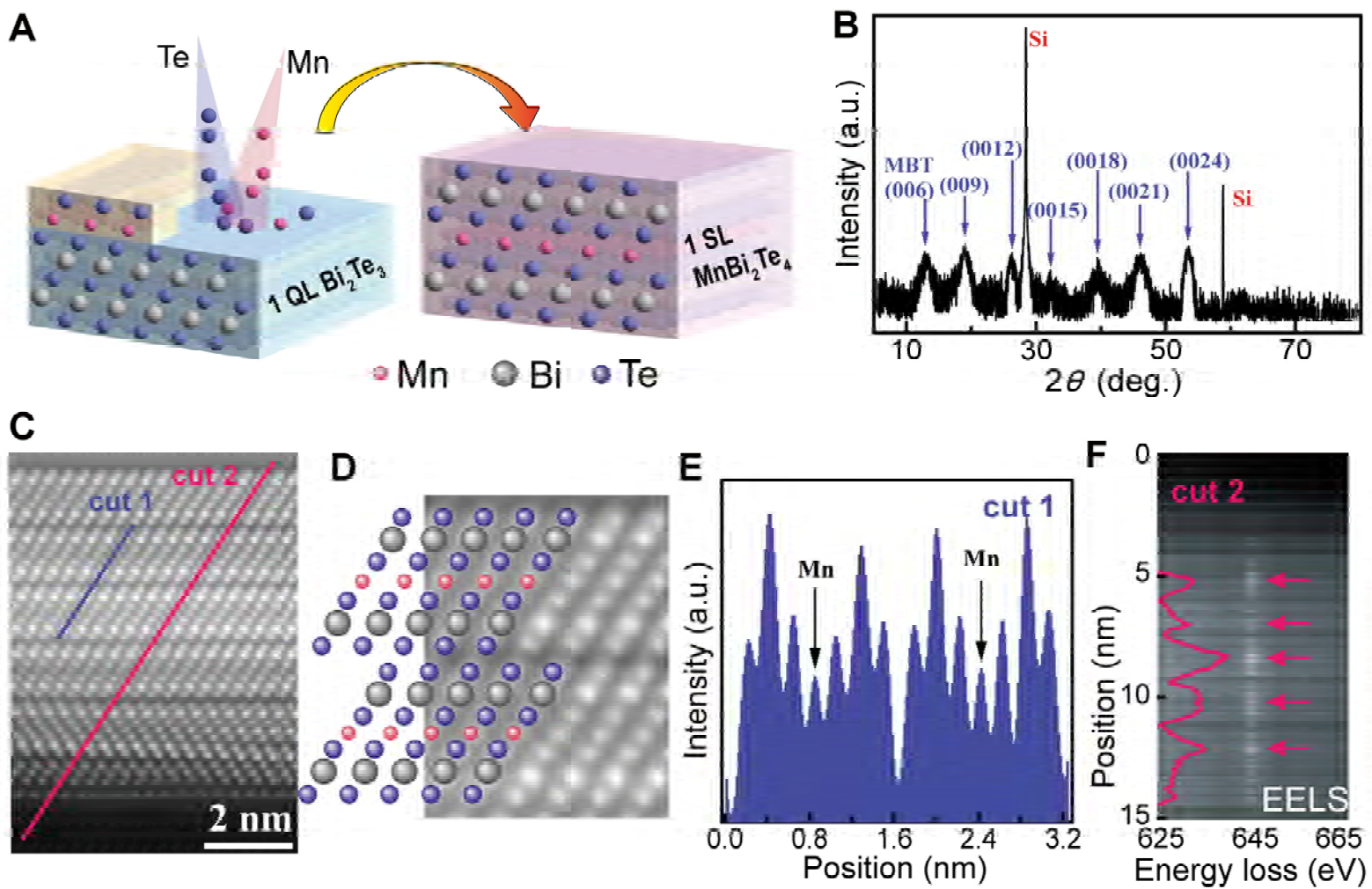

**Figure 2**

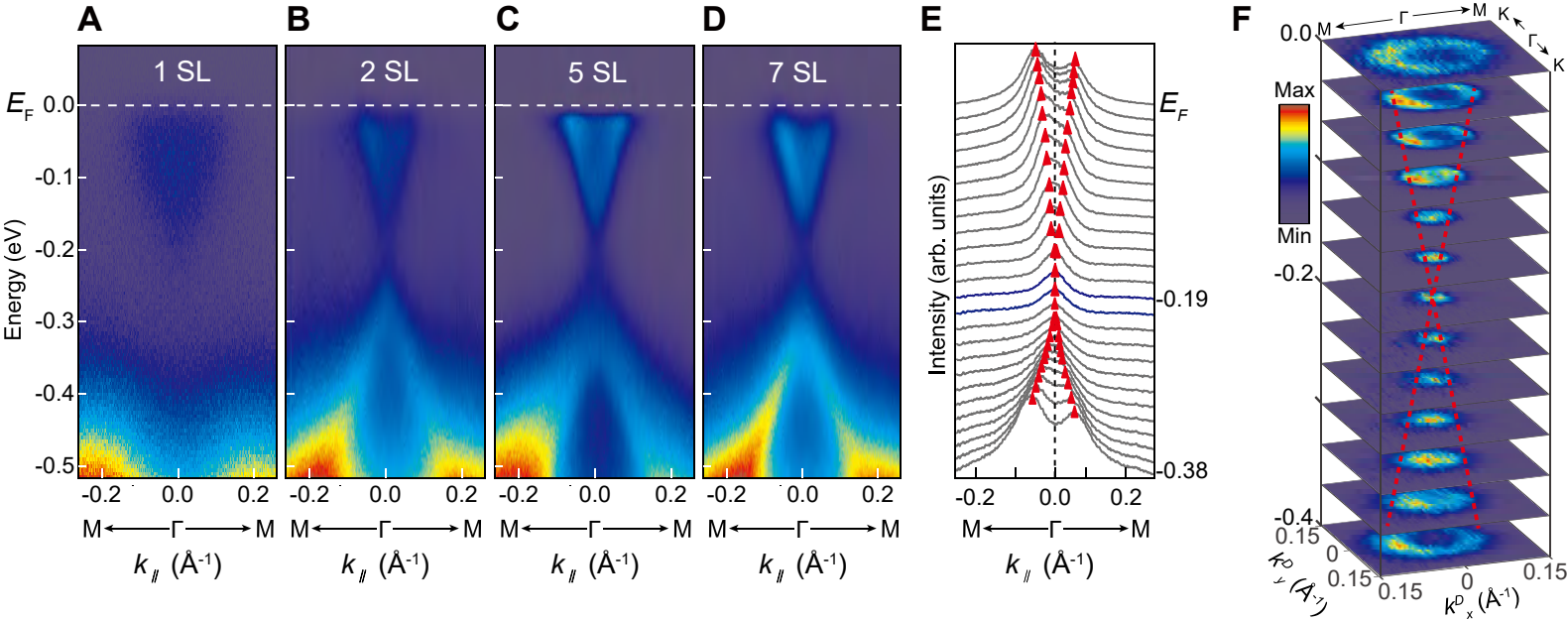

# Figure 3

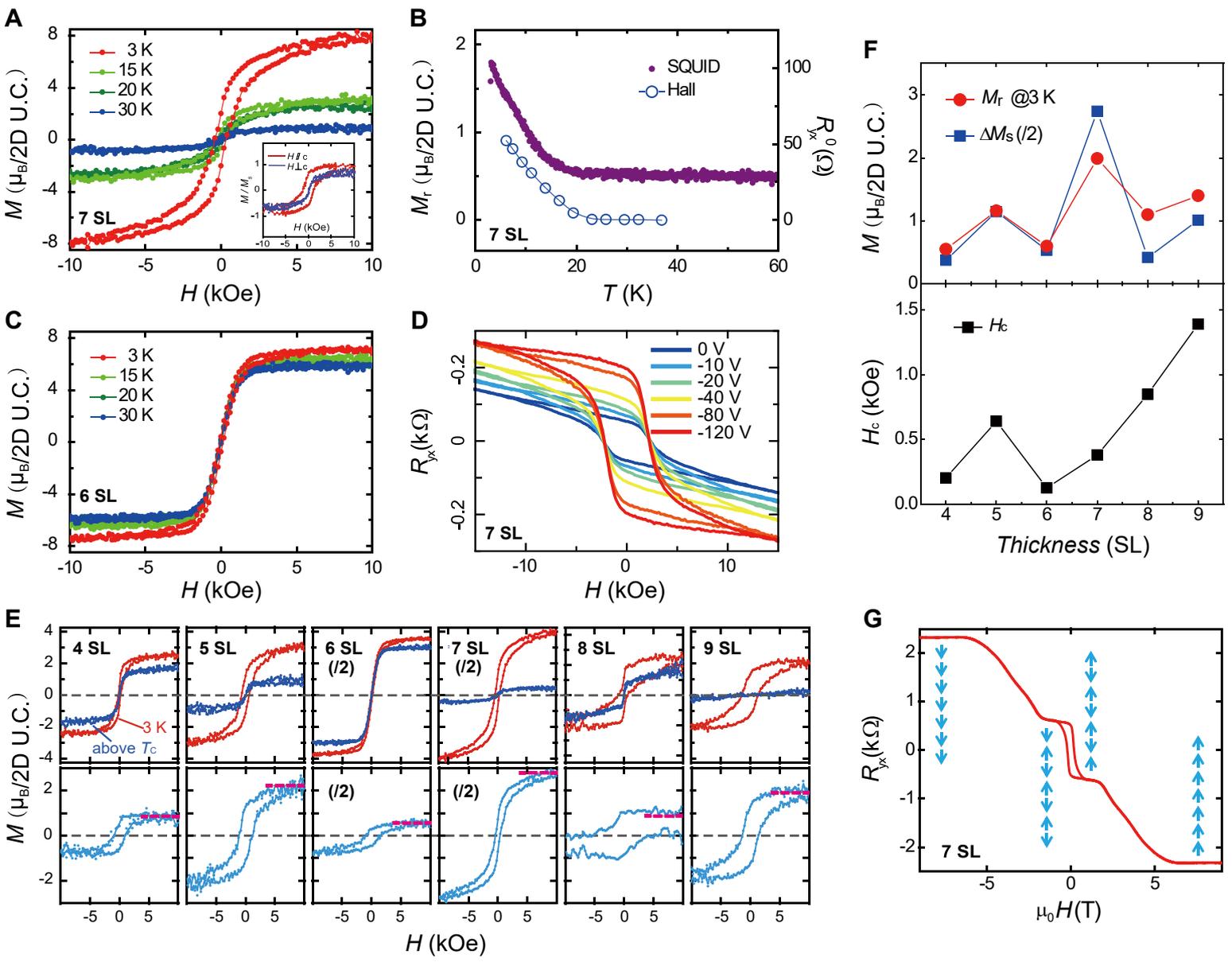

# Figure 4

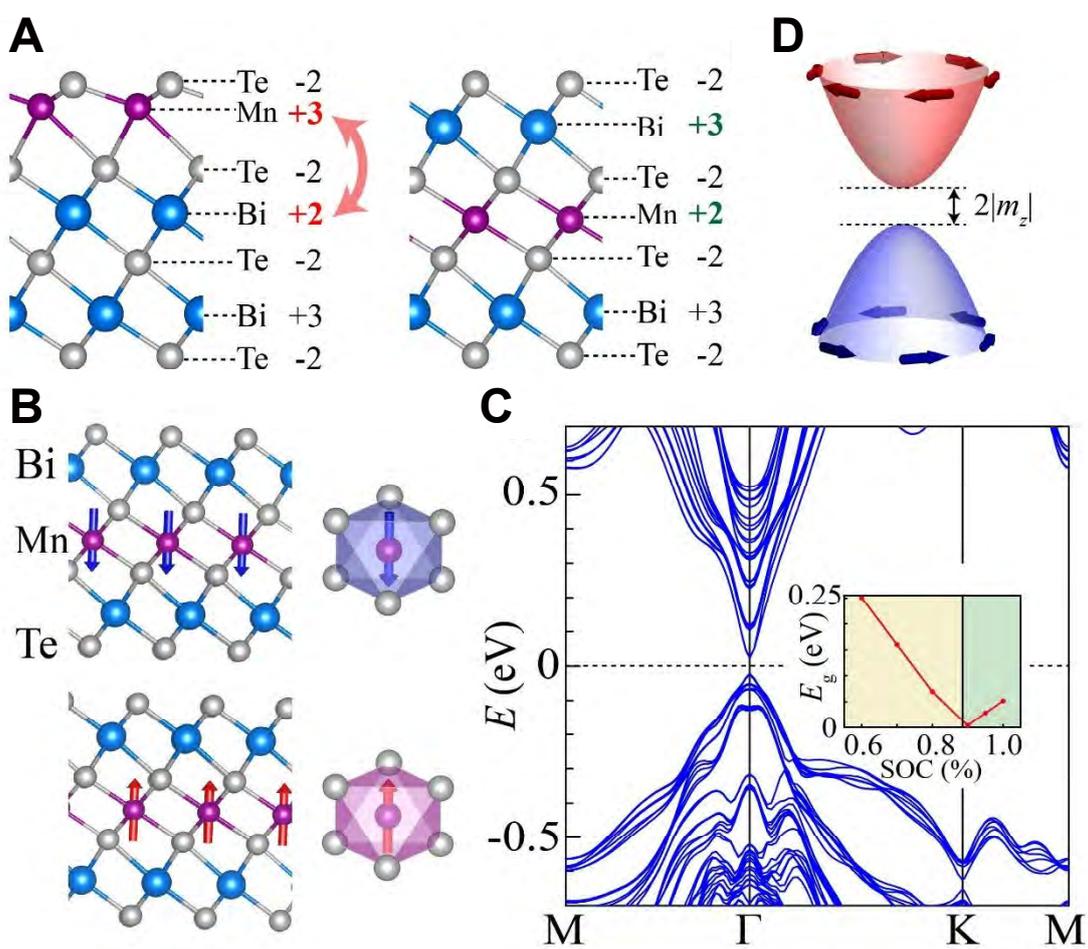